\documentclass[a4paper]{article}
\usepackage{amsmath}
\usepackage{amssymb}
\usepackage{amsfonts}
\usepackage{epsfig}
\usepackage{dcolumn}
\title{Invariant star-products on $S^2$ and \\the canonical trace}
\author{ Keizo Matsubara$^{1}$ \and M\aa rten Stenmark$^2$}
\date{}
\begin{document}
\maketitle
\begin{center}
$^1$Institutionen f\"or Teoretisk Fysik, Uppsala Universitet \\
Box 803, S-751 08 Uppsala, Sweden \\[2mm]
$^2$Institutionen f\"or Str\aa lningsvetenskap, Uppsala Universitet \\
Box 535, S-751 21 Uppsala, Sweden \\[2mm]
{\tt Keizo.Matsubara@teorfys.uu.se, Marten.Stenmark@tsl.uu.se}
\end{center}
\begin{abstract}
In the literature there are two different ways of describing an
invariant star product on $S^2$. We show that the products are
actually the same. We also calculate the canonical trace and use
the Fedosov-Nest-Tsygan index theorem to obtain the characteristic
class of this product.
\end{abstract}
\section{Introduction}
The progress in understanding aspects of noncommutative geometry
has been immense during the past ten years. The formal theory of
deformation quantization \cite{Fronsdal,Kontsevich} has led to
many beautiful discoveries and some use has also been found in the
realm of physics \cite{dn,sz,ars1}, i.e. low energy effective
theories for strings etc.

The most famous deformation quantization is the one which gives
the Moyal product. All other deformations can be identified
locally with this deformation, see for instance \cite{Roche}.

The important interplay between topology and geometry in the
subject of deformation quantization is made explicit in the
theorem by Fedosov, Nest and Tsygan \cite{Fed,Nest}. This theorem
connects the canonical trace to a topological index of the
noncommutative manifold.

This paper has two parts. In the first we investigate the two
invariant star-products on $S^2$ \cite{jap,Alekseev} and show that
they are actually the same. In the last section we consider the
canonical trace. We construct a local isomorphism connecting our
star-product on the sphere and the Moyal product. We use this
isomorphism to calculate the canonical trace of the product on
$S^2$. This gives us the opportunity to calculate the
characteristic class for the star-product on $S^2$, using the
theorem by Fedosov, Nest and Tsygan.

\section{The Star Product}
There are two descriptions of invariant star-products on $S^2$
present in the literature today. These are presented in the papers
\cite{jap,Alekseev}. The main point of this section will be to
prove that the products described are actually the same.

We begin by investigating the star product on the sphere which is
defined in \cite{jap}, and is given by,
\begin{eqnarray}\label{star}
f\star g = fg +\sum_{n=1}^{\infty}C_n(\frac{\hbar}{r})J^{a_1
b_1}\ldots J^{a_n b_n}\partial_{a_1}\ldots
\partial_{a_n}f \partial_{b_1}\ldots \partial_{b_n}g, \nonumber\\
\qquad
\end{eqnarray}
where
\begin{equation}
C_n(\frac{\hbar}{r})=\frac{(\frac{\hbar}{r})^n}{n!(1-\frac{\hbar}{r})(1-2\frac{\hbar}{r})\cdots(1-(n-1)\frac{\hbar}{r})},
\end{equation}
and
\begin{equation}
J^{ab}=r^2\delta_{ab} - x_ax_b + ir\epsilon_{abc}x_c.
\end{equation}
The star product is defined on $\mathbb{R}^3\backslash
\{\mathbf{0}\}$, but can be restricted to two-spheres centered at
the origin since $f(r^2)\star g(\mathbf{x})=g(\mathbf{x})\star
f(r^2)=f(r^2)g(\mathbf{x})$ \cite{jap}, and it is rotation
invariant since $J^{ab}$ is a covariant 2-tensor.

Now, let us turn to the product of the type that can be found in
\cite{Alekseev}. Left invariant vector fields on SU(2) correspond
to elements of the Lie algebra $\mathfrak{su}(2)$. If $X\in
\mathfrak{su}(2)$ we get a vector field acting on functions,
\begin{equation}\label{exp}
  X(f(a))=\frac{\mathrm{d}}{\mathrm{d}t}f(a\exp(tX))\vert_{t=0},
\end{equation}
 $f \in C^{\infty}(\mathrm{SU(2)})$ and $a\in \mathrm{SU(2)}$. If $Z \in \mathfrak{sl}(2,\mathbb{C})$ then,
\begin{equation}\label{comp}
Z(f)=X(f)+iY(f), \quad X,Y \in \mathfrak{su}(2).
\end{equation}

Take $L_a \in \mathfrak{su}(2)$ so that
$[L_a,L_b]=-\epsilon_{abc}L_c$, and define $L_{\pm}=L_1 \pm iL_2$.
The same symbols are used for the Lie algebra elements and the
corresponding vector fields.

We have now described how the operators act on functions on SU(2),
but we want to define operators acting on functions on $S^2$. We
use the fact that $S^2=$ SU(2)/U(1), and let our U(1) be the
subgroup generated by $L_3$. To lift functions on $S^2$ to SU(2)
we use functions on SU(2) satisfying, $f(a)=f(ah):$ $\forall a \in
\mathrm{SU(2)}$, $ \forall h \in \mathrm{U(1)}$. For two such
functions $f$ and $g$ it can be shown that $L_-^nfL_+^ng$ again
has the same property, but $L_{\pm}^nf$ by itself  would not be
defined as a function on $S^2$. This implies that,
\begin{equation}\label{star2}
f\diamond g = fg + \sum_{n=1}^{\infty}C_n(\frac{\hbar}{r})
L_-^nfL_+^ng,
\end{equation}
is a well defined product for functions on $S^2$, where we use the
same symbols for functions on $S^2$ and their lifts.
\newtheorem{equality}{Proposition}
\begin{equality}$\qquad f\diamond g=f \star g, \qquad \forall f,g \in C^{\infty}(S^2)$.
\end{equality}
\emph{Proof.} First we note that, at the unit element, left and
right invariant vector fields agree, and the right invariant
vector fields project down to $S^2$. For $L_a$ the projections to
$S^2$ embedded in $\mathbb{R}^3$ are the well known
$J_a=\epsilon_{abc}x_b\partial_c$. This means that at the point
corresponding to the orbit of the unit, which we call the ``north
pole'', where $x_1=x_2=0$ and $x_3=r$, we can use,
\begin{eqnarray}\label{cor}
L_+=x_3\partial_+ + x_-\partial_3 , \quad L_-=x_3\partial_- +
x_+\partial_3,
\nonumber\\
\mathrm{where}\quad x_-=x_2 - ix_1, \quad x_+=x_2 +ix_1,\nonumber\\
\quad \partial_+= - \partial_{x_2} + i\partial_{x_1}, \quad
\partial_-=-\partial_{x_2} - i\partial_{x_1}.
\end{eqnarray}
It is enough to show that the star products (\ref{star}) and
(\ref{star2}) are the same at one point since they are both
spherically symmetric. To first order the products agree at the
``north pole'' since there
\begin{equation}
J^{ab}\partial_af\partial_bg=L_-fL_+g=\mathcal{L}_-^a\mathcal{L}_+^b\partial_af\partial_bg,
\end{equation}
where we have written the $L_{\pm}$ operators in cartesian
coordinates as $L_{\pm}=\mathcal{L}_{\pm}^a\partial_a$. This means
that $J^{ab}=\mathcal{L}_-^a\mathcal{L}_+^b$ and with our choice
of coordinates (\ref{cor}) it is easy to see using
$\partial_+x_-=\partial_-x_+=0$, that at the ``north pole'' we
have,
\begin{equation}
L_{\pm}^nf=\mathcal{L}_{\pm}^{a_1}\ldots
\mathcal{L}_{\pm}^{a_n}\partial_{a_1}\ldots
\partial_{a_n}f=r^n\partial_{\pm}^nf.
\end{equation}
This means that,
\begin{eqnarray}\label{star=}
L_-^nfL_+^ng =\mathcal{L}_{-}^{a_1}\ldots \mathcal{L}_-^{a_n}\mathcal{L}_{+}^{b_1}\ldots \mathcal{L}_+^{b_n}\partial_{a_1}\ldots \partial_{a_n}f \partial_{b_1}\ldots \partial_{b_n}g \nonumber\\
=J^{a_1 b_1}\ldots J^{a_n b_n}\partial_{a_1}\ldots
\partial_{a_n}f \partial_{b_1}\ldots \partial_{b_n}g=r^{2n}\partial_-^nf\partial_+^ng,
\end{eqnarray}
and when put into the definitions (\ref{star}) and (\ref{star2}),
the proposition follows.$\qquad \square$
\section{The Canonical Trace and the Characteristic class $\theta$}
In this section we will calculate the canonical trace and use the
Fedosov-Nest-Tsygan index theorem \cite{Fed,Nest}, to calculate
the characteristic class of the invariant star-product on $S^2$.
The theorem identifies the canonical trace of the identity to a
topological index of the noncommutative manifold. Before we give
the formulation of the theorem, we will go through the main
definitions of the objects involved.

Given a manifold $M$ and a star-product on it, one can define a
map $f\rightarrow Tr(f)$ s.t. $Tr(f\star g)=Tr(g\star f)$. This
map is cyclic and is called a trace. A distinguished such trace is
the canonical trace defined by the canonical trace density of the
star-product.

 The canonical trace of a star-product may be
defined via the canonical trace of the Moyal product
$\star_m$\cite{Karabegov2},
\begin{equation}
Tr_{can}(f)=\int f\mu_m
\end{equation}
where $\mu_m$ is the formal trace density of the Moyal product as
stated in sec. 3.2. For any other star-product you just replace
the canonical trace density of the Moyal product by that of the
new product. To find the new trace density one does as follows.
Given two star-products, the Moyal product $\star_m$ and some
other product $\star$, they may always be identified on some
neighborhood \cite{Roche}. The two densities can be related due to
the equivalence of the products on the neighborhood. Let
$\mathcal{F}$ denote the equivalence operator of the two
star-products on the neighborhood and call the neighbourhood $U$.
One may then calculate the trace densities, on this neighborhood,
via the equivalence operator as follows,
\begin{equation}
\int_U \mathcal{F} f \mu_{can}=\int_Uf \mu_m
\end{equation}
where $\mu_{can}$ and $\mu_m$ are the trace densities of the two
star-products. Now to calculate the canonical trace globally for
some product other than the Moyal product we would have to connect
the trace densities on an atlas of charts. This may be done since
the densities agree on the overlaps, which make the identification
a well defined procedure. This can however be quite cumbersome.
Due to the symmetries of the product on $S^2$ it will be shown
that the information at one point will be enough for our
calculation.

This gives us some insight into the trace part, but there is more
information needed to formulate the theorem. There are topological
quantities involved in the theorem, since it connects the
canonical trace to certain cohomology classes of the manifold $M$.
These are the three characteristic classes, $\theta(M)$, $c_1(M)$
and Todd$(M)$. The first one, for the symplectic case, belongs to
$\omega/2\pi\hbar+H^2(M)[[\hbar]]$ \cite{Nest} and classifies
star-products up to equivalence. The other two classes are
combined into $\hat{A}(M)=e^{-c_1(M)/2}$Todd$(M)$. Here $c_1(M)$
is the first Chern class of the manifold $M$, with the bundle
structure given by the complex structure induced by the symplectic
form $\omega$. For $S^2$ we have $\hat{A}(S^2)=1$.

Now, let us formulate the index theorem that we will investigate,
the Fedosov-Nest-Tsygan index theorem for a compact symplectic
manifold $M$,
\newtheorem{FNT}{Theorem}
\begin{FNT}
$Tr_{can}(1)=\int_M e^\theta \hat{A}(M)$.
\end{FNT}

We will calculate the characteristic class $\theta$ of our product
by comparing the trace with the topological index. First we will
find the equivalence operator needed to calculate the trace.

\subsection{Transforming the Star Product} Two star-products
$\star_1$ and $\star_2$ defined on the same manifold $M$ are
called equivalent if there exists an operator
$\mathcal{F}=\sum(\frac{\hbar}{r})^kD_k$, where $D_k$ are
differential operators, such that,
\begin{equation}
f \star_2 g =\mathcal{F}^{-1}(\mathcal{F}f \star_1
\mathcal{F}g),\quad \forall f,g \in C^{\infty}(M).
\end{equation}

In our case we want to look at an operator that does such a
transformation around the ``north pole''. We transform the product
in (\ref{star2}), here called $\star_1$, to a star product,
$\star_2$, which we call the polarized Moyal product and is given
by the following expression,
\begin{equation}
f\star_2g=fg
+\sum_{n=1}^{\infty}\frac{(\frac{\hbar}{r})^n}{n!}r^{2n}\partial_-^nf\partial_+^ng.
\end{equation}
This can be transformed into the real Moyal product by doing one
further trivial transformation using the operator
$\mathcal{F}_2=\exp(\frac{\hbar r}{2}\partial_-\partial_+)$.

$\mathcal{F}$ may be taken to be U(1) invariant, a proof of which
may be found in the appendix. This and the fact that it must
transform the original star product correctly give us some
restrictions on how $\mathcal{F}$ can be written. A general
differential operator could be written,
\begin{equation}
\mathcal{F}=\sum c_{ijkl}x_-^ix_+^j\partial_-^k\partial_+^l,
\end{equation}
where the $c_{ijkl}$ are formal numbers, that is elements of
$\mathbb{C}[[\hbar]]$.
 The U(1) invariance gives us immediately the restriction,
\begin{equation}
i+l\neq j+k \Rightarrow c_{ijkl}=0.
\end{equation}

The total $\mathcal{F}$ will have two types of parameters. Some
will contribute at the ``north pole'' and some will not. To see
this take equation (17) and perform all derivatives. At the
``north pole'' all terms for which the derivatives have not
annihilated precisely enough $x_+$ or $x_-$ will be zero. Below we
will concentrate at coefficients that contribute to $\mathcal{F}$
at the north pole, i.e. those for which the derivatives
annihilated precisely enough $x_+$ or $x_-$ to make the
contribution an $\hbar$-dependent constant. The set of parameters,
that are ignored here, might be needed to extend our $\mathcal{F}$
to a neighborhood of this point, but these values will not be
needed in the calculation of the trace. The value of $\mathcal{F}$
at the ``north pole'' is invariant under the change of these
parameters.

 At the north pole the relation,
\begin{equation}
\mathcal{F}(f\star_2 g)= \mathcal{F}f \star_1 \mathcal{F} g,
\end{equation}
reduces to a form where one sees that the $c_{ijkl}$ coefficients
with both $i\neq0$ and $j\neq0$ do not contribute to the result in
that specific point. Of the remaining coefficients we claim that
only those of the form $c_{0k0k}$ or $c_{k0k0}$ can differ from
zero. This can be seen by putting in suitable functions $f$ and
$g$ on both sides and deriving contradictions.  For instance,
assume that $c_{00kk}$ differs from zero then choose $f=x_+^k$ and
$g=x_-^k$ and derive a contradiction. Then assume we have a
nonzero coefficient of the type $c_{a0bc}$ and use two choices of
functions to derive a contradiction. Take first $f=x_-^bx_+^c$,
$g=x_+^a$ and then $f=x_-^a$, $g=x_+^a$, and similarly for the
remaining possibility $c_{0abc}\neq 0$. All this means that we can
write,
\begin{equation}
\mathcal{F}=\mathcal{F}_++\mathcal{F}_-+\mathrm{irrelevant part},
\end{equation}
where $\mathcal{F}_+$ only contains $c_{0k0k}$ coefficients and
$\mathcal{F}_-$ only contains $c_{k0k0}$ coefficients.Observe that
this only holds at the ``north pole''. Now $\mathcal{F}_+$ can be
rewritten as,
%\begin{eqnarray}
%\sum_{n=0}^{\infty}C_n\partial_-^n(\sum c_{i0kl}x_-^i\partial_-^k\partial_+^lf)\partial_+^n(\sum c_{0jkl}x_+^j\partial_-^k\partial_+^lf)=\nonumber \\
%=\sum c_{00kk}\partial_-^k\partial_+^k(\sum_{n=0}^{\infty}\frac{\hbar^n}{n!}\partial_-^n f\partial_+^ng)
%\end{eqnarray}

\begin{equation}
\mathcal{F}_+=F_+(\frac{\hbar}{r},\frac{-x_+}{2}\partial_+)=\sum_{k=0}^{\infty}(\frac{\hbar}{r})^k(-1)^kP_k^+(\frac{-x_+}{2}\partial_+),
\end{equation}
where $P_k^+$ are polynomials. $F_+$ satisfies the relation,
\begin{equation}\label{n+}
\partial_+^nF_+(\frac{\hbar}{r},\frac{-x_+}{2}\partial_+)=F_+(\frac{\hbar}{r},n + \frac{-x_+}{2}\partial_+)\partial_+^n.
\end{equation}
We have a corresponding expression for $\mathcal{F}_-$, just
exchange $+$ with $-$ indices everywhere
\\

We must take $P_0^+(z)+P_0^-(z)=1$, and $P_n^+(0)+P_n^-(0)=0,\quad
\forall n:n>0$, to satisfy the necessary conditions that
$\mathcal{F}(f)$ is equal to $f$ to zeroth order in $\hbar/r$ and
that $\mathcal{F}(1)=1$. This implies that at the ``north pole''
$\mathcal{F}^{-1}=1$. Furthermore we have that
$\partial_\mp^n\mathcal{F}_\pm f=\partial_\mp^nf $. This means
that for the complete expression we get,
\begin{equation}
\mathcal{F}^{-1}(\mathcal{F}f \star_1 \mathcal{F}g)=fg
+\sum_{n=1}^{\infty}C_n(\frac{\hbar}{r})F(\frac{\hbar}{r},n)r^{2n}\partial_-^nf\partial_+^ng,
\end{equation}
where $F=F_++F_-$. Now we see that if we find a function $F$ such
that
\begin{equation}\label{Fn}
F(\frac{\hbar}{r},n)=(1-\frac{\hbar}{r})(1-2\frac{\hbar}{r})\ldots(1-(n-1)\frac{\hbar}{r}),\quad\
n\in \mathbb{Z}^+,
\end{equation}
we will have the wanted product, $\star_2$. This $F$ will have the
same form as $F_+$ but we have $P_k=P_k^++P_k^-$. We can now
formulate the following,

\newtheorem{recursion}[equality]{Proposition}

\begin{recursion} The recursion relation $P_k(z)=(z-1)P_{k-1}(z-1)+P_k(z-1)$,
with $P_0(z)=1$ and $P_k(0)=0,\quad \forall n:n>0$ uniquely
defines $F$ so that it satisfies the relation \emph{(\ref{Fn})}.
\end{recursion}
\emph{Proof.} The function is uniquely specified due to the
uniqueness of polynomials specified at enough points. This can be
seen by looking at the problem combinatorially.  Each polynomial
$P_k(z)$ is specified at enough points to make it unique. The
proof of the recursion relation is by induction. Restrict the
variable to $z=n\in \mathbb{Z}^+$. We first note that the theorem
is valid when $n=1$ which immediately follows from the recursion
relation and the specified values of the polynomials. Now assume
that the theorem is valid for all $n\leq m$. We then look at
$z=m+1$ and use $P_0(z)=1$ to write,
\begin{eqnarray}
&&F(\frac{\hbar}{r},m+1)=1+\sum_{k=1}^{\infty}(\frac{\hbar}{r})^k(-1)^kP_k(m+1)=\nonumber\\&&=1+\sum_{k=1}^{\infty}(\frac{\hbar}{r})^k(-1)^k(mP_{k-1}(m)+P_k(m))\nonumber\\&&=F(\frac{\hbar}{r},m)(1-m\frac{\hbar}{r})=(1-\frac{\hbar}{r})(1-2\frac{\hbar}{r})\ldots(1-m\frac{\hbar}{r}).
\end{eqnarray}
This induction step implies that the relation (\ref{Fn}) holds for
all $n\in\mathbb{Z}^+$, since we know it to be valid for
$n=1$.$\qquad\square$
\\
We can also give the function $F$ in closed form,
\begin{equation}
F(\frac{\hbar}{r},z)=(\frac{\hbar}{r})^{z-1}\frac{\Gamma(\frac{r}{\hbar})}{\Gamma(\frac{r}{\hbar}
-(z-1))}
\end{equation}
as can be seen by a simple calculation where one rewrites the
Gamma functions to obtain (22). It can also be shown to fulfill
the form given in (19).
\subsection{The Canonical Trace Density}

We now take a look at the trace density and calculate it for our
star product. Assume that we have a $2$-dimensional symplectic
manifold $M$ with symplectic 2-form $\omega$. Given this data one
can define the canonical trace density on the Moyal product to be,
\begin{equation}
\mu_{m}=\frac{\omega}{2\pi\hbar},
\end{equation}
where,
\begin{equation}
\int_M\omega=2\pi r.
\end{equation}
The canonical trace density for any other star product on our
manifold can now be calculated by using the pullback of the
canonical trace density of the Moyal product using the operator
$\mathcal{F}$. The canonical trace densities of the Moyal product
and the polarized Moyal products are the same, so we could use our
operator instead of the one that transforms to the real Moyal
product.
 According to theorem \cite{Karabegov2}, any two trace densities
of the same star product can only differ by a multiple of a formal
number. That is, if $\mu_1$ and $\mu_2$ are two different trace
densities of the same star product, $\mu_1=c(\hbar)\mu_2$.

\newtheorem{defi}{Definition}
\begin{defi}
A star product is called strongly closed if $\omega$ is a trace
density.
\end{defi}
This together with the above given observations imply that for a
strongly closed star product $\mu_{can}=c({\hbar})\mu_m$. We also
state the following theorem which can be found in
\cite{Karabegov2}.
\newtheorem{strong}[FNT]{Theorem}
\begin{strong}
If a Lie group $G$ acts transitively on $M$ by symplectomorphism
in such a way that the corresponding shifts in the algebra are
automorphisms i.e. if $G$ is a symmetry group of the star product,
then the star product is strongly closed.
\end{strong}
In our case since $\star_1$ is spherically symmetric and SO(3) is
a symmetry group of our star product we know that it is strongly
closed. So the canonical
 trace density will differ everywhere from the canonical Moyal trace density by a
 formal number $c(\hbar)$.

 To completely define the canonical trace density of our product,
 $\star_1$,
  it would therefore be enough to find the value of $c(\hbar)$. Due to the spherical
  symmetry this can be done by looking at one point, which gives us the relation
\begin{eqnarray}
c(\hbar)\int_U \mathcal{F}\delta_0 \mu_m&=& \int_U \delta_0
\mu_m \\
&\Rightarrow&\nonumber\\
 c(\hbar)&=& \left(\frac{\int_U
\delta_0 \mu_m}{ \int_U \mathcal{F}\delta_0 \mu_m}
\right)\nonumber\\
&=&F(\frac{\hbar}{r},-1)^{-1}
\end{eqnarray}
$c(\hbar)$ can now be calculated using the following proposition.

\newtheorem{F(-1)}[equality]{Proposition}
\begin{F(-1)}$F(\hbar/r,-1)=(1+\hbar/r)^{-1},\qquad \hbar/r\neq-1$
\end{F(-1)}
\emph{Proof.} From the recursion relation it immediately follows that $\forall k:$ $P_k(-1)=1$
when this is put into the definition of $F(\hbar/r,-1)$ we get the Taylor expansion
of $(1+\hbar/r)^{-1}$, which converges to the function except when $\hbar/r=-1$ and
this proves the theorem.$\qquad \square$\\
This result means that,
\begin{equation}
\mathrm{Tr}_{can}(1)=\left(1+
\frac{\hbar}{r}\right)\int\mu_{m}=1+\frac{r}{\hbar}.
\end{equation}
\newtheorem{car}[equality]{Proposition}
\begin{car}The characteristic class of the invariant
star-product on $S^2$ is $\theta=\omega/2\pi\hbar+c_1(S^2)/2$.
\end{car}
\emph{Proof.} We know that $\theta(S^2)=\omega/2\pi\hbar+\rho$
with $\rho\in H^2(S^2)[[\hbar]]$ \cite{Roche}, and that
$\hat{A}(S^2)=1$ as stated in the introduction to this section.
For compact two dimensional manifolds with a complex structure we
know that the integral of $c_1$ equals the Euler character. If we
choose a complex structure $J$, as in \cite{Nest}, such that
$\omega (Jx,y)$ is positive definite, one has $\chi=2$. This
information is all that is needed for our calculation.

Using the results above we calculate,
\begin{eqnarray}
\int_{S^2} e^\theta \hat{A}&=&\frac{r}{\hbar}+\int_{S^2}\rho,
\end{eqnarray}
so that we get, from the index theorem, the value $\int\rho=1$ for
our product. Now the choice $\rho=c_1(S^2)/2$ gives the desired
result. Here one uses that $H^2(S^2, \mathbb{R})=\mathbb{R}$.
Hence, the integral of $\rho$ is sufficient to identify $\rho$.$\qquad\square$\\

\subsubsection*{Acknowledgements} Our sincere thanks to
Anton Yu.\!\! Alekseev without whom this project would never have
been accomplished.

\section*{Appendix}
We argue that we can choose the map between $\star_1$ and
$\star_2$ to be U(1) invariant. To prove this we need some
definitions.

Let the functions $a$, $b$, $f\in C^\infty (U)$, be such that $a$
is holomorphic and $b$ antiholomorphic. A star-product will be
called a star-product with separation of variables if for any open
subset $U\subset M$ of a symplectic manifold $M$ it holds that
$a\star f=af$ and $f\star b=fb$.

Furthermore let the star-product algebra $\mathcal{A}$ be the
algebra of functions written as formal series in $\hbar$, given by
the product $\star$. Let $C^k(\mathcal{A},\mathcal{A})$ be the
space of mappings from, the space of $k$-multi-differential
operators on $M$ $\mathcal{A}\otimes ...\otimes\mathcal{A}$, to
$\mathcal{A}$.

Between $C^k(\mathcal{A,A})$ and $C^{k+1}(\mathcal{A,A})$, define
the Hochschild coboundary operator $\tilde{b}$, such that,
\begin{eqnarray}
(\tilde{b}c)(u_0,...,u_k)&=&u_0\star
c(u_1,...,u_k)+\sum_{i=0}^{k-1}(-1)^{i+1}c(u_0,...,u_i\star
u_{i+1,...,u_k})\nonumber
\\
&&+(-1)^{k+1}c(u_0,...,u_{k-1})\star u_k,
\end{eqnarray}
for $u_0,...,u_k\in\mathcal{A}$ and $c\in C^k(\mathcal{A,A})$. The
Hochschild cohomology groups are the cohomology groups of this
complex.

Now consider the specific products of section 3. First note that
deformations of K\"ahler forms $\omega_{*_i}$ are then in 1-1
correspondence with certain star-products $\star_i$ with
separation of variables\cite{Karabegov1}. Hence to each of the
products there is a specific K\"ahler forms. Since
$\omega_{\star_2}$ belongs to the same cohomology class as
$\omega_{\star_1}$ on a contractible subset, we know, due to
Moser´s argument, that for some smooth family of 1-forms
$\beta_t\in\Omega^1(S^2)$ there exists a family of diffeomorphisms
$\phi_t$ such that for $\omega_t=\omega_0+d\beta_t$ one has
$\phi^*_t\omega_{\star_t}=\omega_{\star_1}$\cite{Moser}. This
gives us $\phi^*_{t=1}\omega_{\star_2}=\omega_{\star_1}$. Hence we
may identify $\omega_{*_1}$ with $\omega_{*_2}$, using $\phi_{t}$.

This will induce an isomorphism between the algebras
$\mathcal{A}_1$ and $\mathcal{A}_2$, were the index indicates
which star product it corresponds to, since there is a 1-1
correspondence between the deformation quantization and the
deformations of K\"ahler forms.

Writing the isomorphism $D_t$ one has $f\star_t
g=D_t^{-1}(D_t(f)\star_1 D_t(g))$ which infinitesimally may be
written as $f\star_t g=f\star_1 g+\epsilon\alpha_t (f,g)$ where
$\alpha_t$ is a map from $\mathcal{A}\otimes\mathcal{A}$ to
$\mathcal{A}$.

Now, since demanding $\star_t$ to be associative is equivalent to
demanding that \~{b}$\alpha_t=0$, where \~{b} is the Hochschild
coboundary operator we know that we may write $\alpha$ as \~{b}$c$
for some $c:\mathcal{A}\rightarrow \mathcal{A}$, due to the
vanishing of Hochschild cohomology groups on contractible
subsets\cite{Xu}. Hence $\alpha_t$ may be written,
\begin{equation}
\alpha_t (f,g)=-c_t (f\star_1 g)+c_t (f)\star_1 g+f\star_1 c_t (g)
\end{equation}

We now know that there exists an isomorphism $D_t$ and also its
infinitesimal properties given by the Hochschild coboundary
operator. Now we may average over $c$'s using the  U(1)-action and
produce a U(1) invariant isomorphism $D$ which may be integrated
and thus extended to a non infinitesimal isomorphism. Hence we may
choose $\mathcal{F}$ to be U(1) invariant.

\end{document}